# On Angular Momentum and Magnetic Moment in Many-Electron System

Yuri Kornyushin

*Maître Jean Brunschvig Research Unit,*
*Chalet Shalva, Randogne,*
*CH-3975, Switzerland*

Simple explanation of coexistence of zero angular momentum and non-zero magnetic moment in many-electron system is discussed in this short note from the point of view of statistical physics.

In this short note I would like to comment a problem of statistical behavior of many-electron system in external applied magnetic field [1]. It is stated in [1] that in classical and quantum statistics application of external magnetic field does not lead to the changes in magnetic properties of a sample (Bohr-van Leeuwen theorem). This means that it is stated in [1] that classical and quantum statistics could not describe phenomenon of diamagnetism in many-electron system. It is stated also in [1] that many-electron system should conserve initial zero angular momentum after magnetic field is applied. Taking into account this conservation makes it possible to describe the diamagnetism phenomenon in many-electron system [1]. As is well-known [2, 3], the angular momentum of an individual electron, $\mathbf{L} = m\mathbf{r}\times\mathbf{v}$, whereas its magnetic moment, $\boldsymbol{\mu} = -(ge/2c)\mathbf{r}\times\mathbf{v}$. Here $m$ is the electron effective mass, $\mathbf{r}$ is electron coordinate vector, $\mathbf{v}$ is electron velocity, $g$ is Landé $g$-factor, $e$ is the elementary charge, and $c$ is the speed of light in vacuum. It looks like electron magnetic moment is proportional to its angular momentum. It is so for an individual electron. It should be noted here that the individual electron angular momentum is proportional to its effective mass, whereas the individual electron magnetic moment does not depend on its effective mass at all.

    It is well-known that many-electron system in a solid state can be described statistically, in particular, as many electrons, having different effective masses [2]. It is well-known that the effective masses can be of different values, and that the effective mass can be negative due to circumstances [2]. When there is a single effective mass in a system, for such a system magnetic moment of a system is proportional to its angular momentum. From this follows that for such a system with zero angular momentum, its magnetic moment has to be zero also.

    In real systems, as is well-known, in energy regions far away from minimum effective mass can be negative and of a comparatively very large value [2]. In metals, in particular, the majority of the delocalized electrons are in energy regions far from minimum.

So it looks rather a common case for delocalized electrons in metals that they have total zero angular momentum (the sum of the individual angular momentums with different positive and negative effective masses), while the total magnetic moment of a sample (the sum of the addendums not depending on the effective masses) can be of essential value.

The specified here explanation of the coexistence of zero total angular momentum and non-zero total magnetic moment in many-electron system seems rather simple and relevant.